# Clustering Based Lifetime Maximizing Aggregation Tree for Wireless Sensor Networks

Deepali Virmani
CSE BPIT, GGSIPU
New Delhi, India
deepalivirmani@gmail.com

Satbir Jain
CSE, NSIT,
New Delhi, India
Jain_satbir @yahoo.co.in

*Abstract*— **Energy efficiency is the most important issue in all facets of wireless sensor networks (WSNs) operations because of the limited and non-replenish able energy supply. Data aggregation mechanism is one of the possible solutions to prolong the life time of sensor nodes and on the other hand it also helps in eliminating the data redundancy and improving the accuracy of information gathering, is essential for WSNs. In this paper we propose a Clustering based lifetime maximizing aggregation tree (CLMAT) in which we create aggregation tree which aim to reduce energy consumption, minimizing the distance traversed and minimizing the cost in terms of energy consumption. In CLMAT the node having maximum available energy is used as parent node/ aggregator node. We concluded with the best possible aggregation tree minimizing energy utilization, minimizing cost and hence maximizing network lifetime.**

## I. INTRODUCTION

WSNs are communication networks which rely on the collective information provided by sensors but not on any individual sensing report. A node in WSNs, at one specific time may be granted more access to the network than all other nodes if the program objective is still satisfied. Network resources are shared as long as the application performance is not degraded. As sensors are being densely-deployed in WSNs, the detection of a particular stimulus can trigger the response from many nearby nodes. So instead of data being send by all the nodes collectively, the data is aggregated from neighboring sources locally to remove any redundancy and produce a more concrete reading [7][8]. None of the previous works have considered the three most important factors (energy, distance, and cost) simultaneously. We focus on constructing a data aggregation tree among any given set of source nodes present in the initial network keeping all the three factors in view. Moreover, the tree is structured in a way that can enhance the lifetime of the sensor network. Reference [9] suggests that extending node lifetime is equivalent to increasing the amount of information gathered by the tree root when the data rate is not time-varying. We consider a network of randomly- deployed sensor nodes in which each node has an identical transmission range. An event that triggers the sensors around it occurs at random in the network. All the node present in the network send their data to a root node in the tree such that minimum distance is been traversed, and minimizing the cost in terms of energy consumption and also the energy consumed is less. Thus taking all the above factors in mind a final tree is been selected which leads to the maximization of network lifetime, minimizing energy consumption, minimizing distance traversed and minimization of cost.

## II. LITERATURE AND REVIEW

### A. Data Aggregation

The data aggregation is a technique used to solve the implosion and overlap problems in data centric routing. Data coming from multiple sensor nodes are aggregated as if they are about the same attribute of the phenomenon when they reach the same routing node on the way back to the sink. Data aggregation is a widely used technique in wireless sensor networks. The security issues, data confidentiality and integrity, in data aggregation become vital when the sensor network is deployed in a hostile environment.

### B. Literature Review

Our work bears some resemblance to other research efforts in the literature. In fact, a number of recent works has begun to consider collaborating nearby sensor nodes by the use of a data aggregation tree/cluster. Such tree/cluster provides event sources with a mechanism to refine their readings, so that only a minimum amount of energy is required to deliver the information to the user. In this section, we provide a summary of these construction techniques.

1) *Energy-Aware Data Aggregation Tree (EADAT)*
The work in [1] attempts to construct a tree rooted at a base station and spanned all network nodes by extensive use of timers. Motivated by the fact that only non-leaf nodes in the tree are aggregating and relaying traffic, radios of all leaf nodes are turned off for immediate energy savings. The nodes with higher residual energy have a higher chance to become non leaf tree; thereby enhancing the likelihood of turning off lower-energy leafs. The algorithm requires a given tree root (base station) to initially broadcast a control message and start the tree construction. Each node upon receiving this message for the first time starts a timer that expires in time duration inversely proportional to its residual energy. A timer is refreshed if a node receives a message during the count down. After the timer expires, the node broadcasts a similar control message indicating its willingness to be a parent in the tree.

2) *Maximum Lifetime Data Aggregation*
Reference [2] attempts to find a schedule of various directed trees, subject to the requirement that the number of rounds during which a base station can aggregate information from all the nodes via these trees is maximized. The protocol assumes that nodes are aware of every other's positions and have the abilities to directly reach any other sensor (including the base station) in the network. Such a Maximum Lifetime Data Aggregation (MLDA) problem is approached by coordinating the radio ranges and data aggregating agents of various nodes in a way that the resultant flow of traffic towards the base station maximizes the system lifetime.

3) *Minimum-Cost Convoy Tree*
More recent work has begun to consider collaborating nearby sensor nodes to generate a more concrete report of the object being traced. Such issue has been recognized by which further provides a dynamic convoy tree-based collaboration (DCTC) framework for tracking a mobile target. The root can dynamically collect and refine the readings gathered from various tree nodes. The challenge of their work lies on finding a sequence of minimum cost trees, so called minimum-cost convoy tree sequence, whose coverage on the moving object is above a certain threshold. The tree they have considered is the one that has the root being closest to the target. Furthermore, all other nodes are arranged in a way that the cost of sending a packet via some nodes to this root is minimized [3].

4) *Spanning Tree over Area-Dominating Set*
The smallest subset of nodes that covers the monitoring area is referred to as the area-dominating set. The authors in this paper suggest the use of a spanning tree, induced by the initial interest flooding over the area-dominating set, for aggregating reply messages from various event sources. The sink where the interest is originated from is the root of the spanning tree. As with DCTC, the authors did not consider node's residual energy in the tree construction. The result can be a reduction in the node's lifetime and the amount of information collected by the tree root [4].

5) *Balanced Converge cast Tree*
The work in [5] addresses the problem of converge cast (many-to-one) for data aggregation. A tree that is rooted at the base station is constructed so that the link cost from each node to the base station is minimized. The authors further improve the design by balancing the tree during the construction, thereby enhancing the likelihood of simultaneous aggregation and reducing the latency for converge cast. Furthermore, two Code Division Multiple Access (CDMA) codes are allocated to nodes for collision-free transmissions towards the base station.

6) *The Energy-Aware Spanning Tree (E-Span)*
Before starting to describe our LPT algorithm, we outline the basic spanning tree protocol [6] followed by presenting an energy-aware variant of it, namely E-Span. We believe that E-Span shall provide some insights on how different event sources should be arranged in the lifetime preserving tree and is likely to satisfy our objectives for only a few participating source nodes.

III. CLMAT: PROPOSED AGGREGATION TREE

In all the above mentioned work some has taken into consideration energy, some has taken cost, some has taken distance, but none of them has considered all these factors together. So keeping into mind the energy conservation i.e. to reduce the energy consumed we have also considered minimum cost consumption, shortest distance covered and providing backup for the existing sink.

Proposed key features of aggregation tree are:
1) Enhance network lifetime
2) Minimize Energy Consumption
3) Minimize Distance
4) Minimize Cost Consumption

The branch energy is calculated as follows:
$brE_{x,y,A} = \min \{e_i\}$   (1)
$i \in A_x, i \neq x$

where
$brE_{x,y,A}$ : Energy of branch $A_x$ leafed at node x and rooted at node y, $A \in P_{x,y}$
$I_z$ : Set of nodes in given tree rooted at node z
$TEI_z$ : Energy of tree rooted at node z
$P_{x,y}$ : path b/w node x and y.

The tree energy is calculated as follows:
$TEI_z = \min \{e_j\}$    (2)
$j \in I_z, j \neq x$
where
$TEI_z$ : Energy of tree rootes at node
$I_z$ : Set of nodes in given tree rooted at node z

The cost [10] is calculated as follows:

$EC_{(m,n)} = (e_{mn}) \tilde{E}_m^{-1} + (e_{nm}) \tilde{E}_n^{-1}$    (3)
where
$EC_{(m,n)}$ : Energy cost for transmitting a packet from node m to node n
$e_{mn}$ : Transmission Energy required for node n to transmit a bit to its neighboring node n
$\tilde{E}_m$ : Residual energy of node m

The distance is calculated as follows:
$D_y = \Sigma d_{y,a}$

where
$D_y$ : Total distance of tree rooted at y

$d_{y,a}$ : Distance from root y to node a

### A. Proposed Algorithm
1.) Create the graph

Set 'n' to number of sensor nodes.
   Graph()
1. for i to (n-1)
2. for j to (n-1)
3. if i equals j
4. distance[i][j]=0
5. else
6. distance[i][j]=infinity

2.) Checking existing edge

   edgeExists()
1. for i to (n-1)
2. for j to (n-1)
3. if (distance[i][j]!=0) and (distance[i][j]!=infinity)
4. Return true;
5. Return false;

3.) Get Index for vertex name

   getIndex(*vname)
1. for i to (n-1)
2. if(strcmp(vertices[i],vname)==0)
3. return i;
4. else
5. return -1;

4.) Add Vertex to Graph

   addVertex()
1. i=getIndex(vname)
2. if i!=-1
3. Vertex already exists.
4. return
5. strcpy(vertices[n],vname)
6. value of 'n' is incremented by 1

5.) Add Edge To Graph

   addEdge()
1. if n=0
2. No vertex exists.
3. return
4. while(True)
5. Source vertex v1 is entered
6. i1=getIndex(v1)
7. if(i1=-1)
8. source vertex does not exist.
9. else
10. Enter energy_vertex[i1]
11. Destination vertex v2 is entered
12. i2=getIndex(v2)

13. if(i2=-1)
14. Destination vertex does not exist.
15. else
16. Enter energy_vertex[i2]
17. break
18. if(energy_vertex[i1]>energy_vertex[i2])
19. energy_edge[i1][i2]=energy_vertex[i2]
20. else
21. energy_edge[i1][i2]=energy_vertex[i1]
22. Enter distance[i1][i2]

*6. ) Display Graph*

display()
1. if(n=0)
2. Graph does not exist.
3. return
4. for i=0 to (n-1)
5. vertices[i]
6. if edgeExists()
7. for i=0 to (n-1)
8. for j=0 to (n-1)
9. if((distance[i][j]!=0)and(distance[i][j]!=infinit))
10. vertices[i]->vertices[j]-distance[i][j]--energy_edge[i][j]

*7.) Find Shortest Path & Cost of path traversed & Energy of tree*

findShortestPathCostEnergy()
1. if(n=0)
2. Graph does not exist.
3. return
4. while(True)
5. src=getIndex(source)
6. if(src=-1)
7. Source vertex does not exist.
8. else
9. break
10. for i=0 to (n-1)
11. Distance[i]=distance[src][i]
12. Final[i]=0
13. Final[src]=1
14. for i=0 to (n-1)
15. for j=0 to (n-1)
16. if(Final[j]=0)
17. v=j
18. break
19. for j=0 to (n-1)
20. if (Final[j]=0) and (Distance[j]<Distance[v])
21. v=j
22. Final[v]=1
23. for w=0 to (n-1)
24. if(Final[w]=0)
25. if(Distance[v]+distance[v][w]<Distance[w])
26. Distance[w]=Distance[v]+distance[v][w]
27. for i=0 to (n-1)
28. for j=0 to n-1
29. if(Distance[j]=infinity)
30. source->vertices[j]=No path
31. else
32. source->vertices[j]=Distance[j]
33. if(energy_vertex[i]>energy_vertex[j])
34. Edge _Energy[i][j]=energy_vertex[j]
35. else
36. Edge_Energy[i][j] = energy_vertex[i]
37. Total_Distance[i] = Total_Distance[i] + Distance[j]
38. a=Edge_Energy[0][1]
38. for i=0 to n-1
39. for j=0 to n-1
40. if(Edge_Energy[i][j]<a)
41. Tree_Energy[i]=Edge_Energy[i][j]
42. Residual_Energy1 = energy_vertex1-Tree_Energy
43. Residual_Energy2 = energy_vertex2 - Tree_Energy
44. cost_edge[n][n]=( energy_vertex1 / Residual_Energy1)+(energy_vertex2 / Residual_Energy2)
45. for i=0 to n-1
46. for j=0 to n-1
47. cost_tree[i]=cost_tree[i] + cost_edge[i][j]

*8.) Comparison of Aggregation Trees*

CompareTrees( )
1. a=Total_Distance[0]
2. for i=0 to n-1
3. if(Total_Distance[i] <a)
4. a= Total_Distance[i];
5. k=i;
6. b=Tree_Energy[k]
7. c=cost_tree[k]
8. ith tree is the final tree with a, b &c as distance, energy & cost
9. If two trees have same parameters, their depth traversal is considered.

*9.) Clustering Based Lifetime Maximizing Aggregation Tree (CLMAT)*

```
    Main()
1.  do
2.  Switch(ch)
3.  if ch=1
4.  addVertex()
5.  break
6.  if ch=2
7.  addEdge()
8.   break
9.   If ch=3
10. display()
11. break
12. if ch=4
13. findShortestPathCostEnergy()
14. break
15. if ch=5
16. CompareTrees()
17. if ch=6
18. exit()
19.  while(ch!=6)
```

A. *Aggregation Trees*

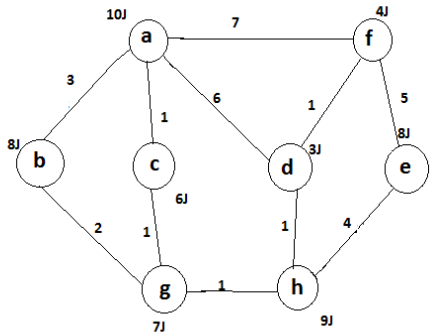

Figure 1. Graph Assumed (Initial Scenario)

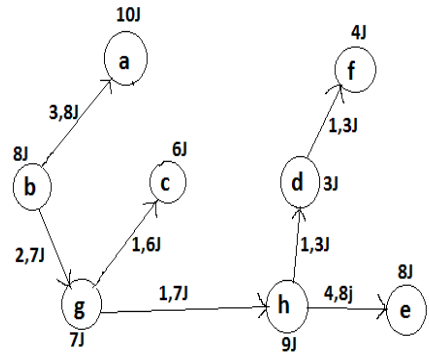

Energy: 3J, Cost: 21.978, Distance: 27
Figure 3. Node B as an Aggregator node

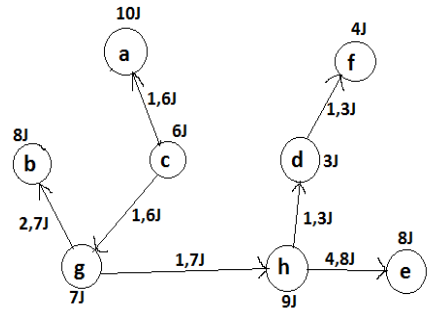

Energy: 3J, Cost: 22.278, Distance: 20
Figure 4. Node C as an Aggregator node

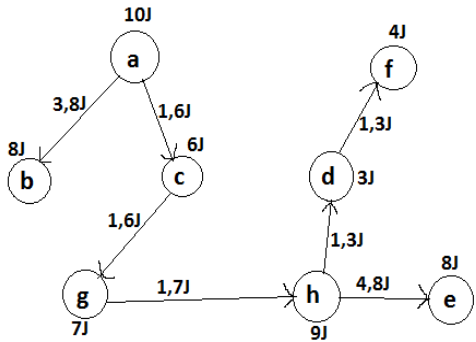

Energy: 3J, Cost: 22.056, Distance: 25
Figure 2. Node A as an Aggregator node

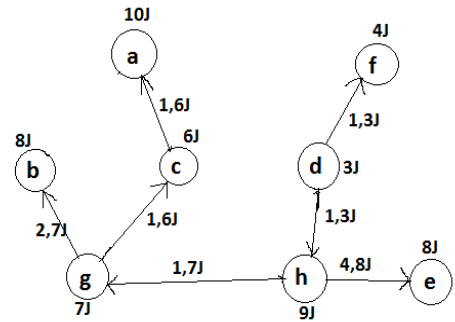

Energy: 3J, Cost: 22.378, Distance: 20
Figure 5. Node D as an Aggregator node

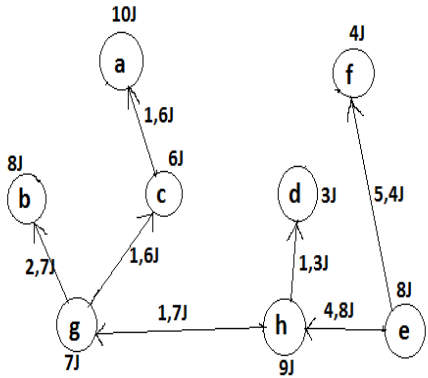

Energy: 3J, Cost: 23.978, Distance: 39
Figure 6. Node E as an Aggregator node

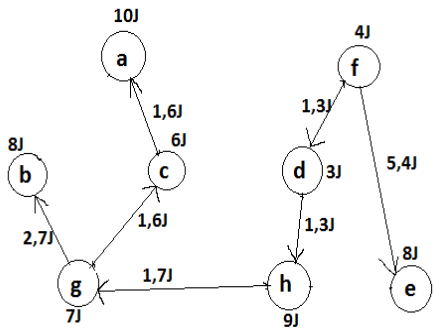

Energy: 3J, Cost: 24.878, Distance: 25
Figure 7. Node F as an Aggregator node

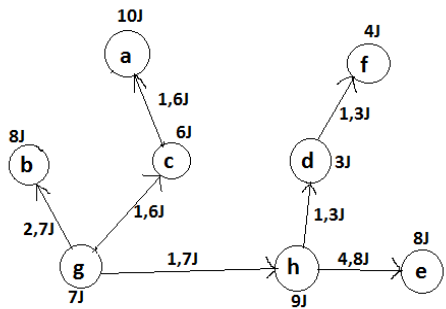

Energy: 3J, Cost: 22.378, Distance: 16
Figure 8. Node G as an Aggregator node

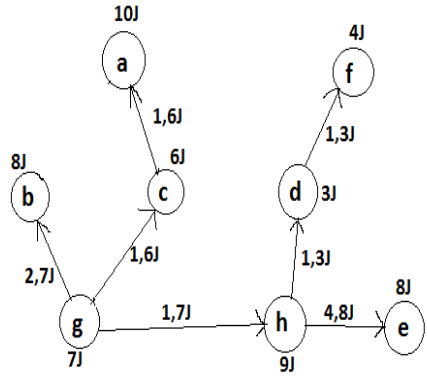

Energy: 3J, Cost: 22.378, Distance: 16
Figure 9. Node H as an Aggregator node

### B. Final Aggregation Tree

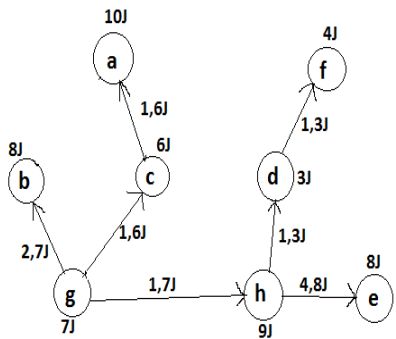

Energy: 3J, Cost: 22.378, Distance: 16
Figure 10. Node H as an Aggregator node

So it is clear by applying the above proposed CLMTA algorithm we able to achieve the best possible aggregation tree. The achieved tree is using minimum energy, traversing minimum distance and cost in terms of energy is minimized for the achieved tree having H as the aggregator node.

## CONCLUSION & FUTURE WORK

In this paper we proposed a clustering based lifetime maximizing aggregation tree (CLMAT) in which creates an aggregation tree which aims to reduce energy consumption, minimize the distance traversed and minimize the cost in terms of energy consumption. In CLMAT the node having maximum available energy is used as parent node/ aggregator node. Final tree produced using the proposed algorithm proves to use minimum energy as well as traverses the minimum distance and uses the minimum cost in terms of energy consumption. Hence by achieving the above mentioned parameters the obtained aggregation tree proves to be the best for enhancing the network lifetime.

## REFERENCES

[1] M.Ding, X.Cheng, and G.Xue, "Aggregation tree construction in sensor networks," in Proc. of IEEE Vehicular Technology Conference (VTC'03), vol. 4, Orlando, FL, pp. 2168-2172, Oct. 2003.

[2] K. Dasgupta, K. Kalpakis, and P. Namjoshi, "An efficient clustering-based heuristic for data gathering and aggregation in sensor networks," in Proc. of IEEE Wireless Communications and Networking Conference (WCNC'03), New Orleans, LA, pp. 1948-1953, Mar. 2003.

[3] W. Zhang and G. Cao, "DCTC: Dynamic convoy tree-based collaboration for target tracking in sensor networks," IEEE Trans. Wireless Commun., vol. 3, no. 5, pp. 1689-1701, Sept. 2004.

[4] J. Carle and D. Simplot-Ryl, "Energy-efficient area monitoring for sensor networks,"IEEE Computer Magazine, vol. 37, no. 2, pp. 40-46, Feb. 2004.

[5] S. Upadhyayula, V. Annamalai, S. K. S. Gupta, "A low-latency and energy-efficient algorithm for converge cast," in Proc. of IEEE Global Telecommunications Conference (GLOBECOM'03), vol. 6, pp. 3525-3530, Dec. 2003.

[6] R. Perlman, "Interconnections: Bridges, routers, switches", and internetworking protocol, 2nd ed., Addison-Wesley Professional Computing Series, Reading, MA, 1999.

[7] I. Nikolaidis, J. J. Harms, and S. Zhou, "On sensor data aggregation with redundancy removal," in Proc. of 22nd Biennial Symposium on Communications, Ontario, CA, May 2004.

[8] A. Boulis, S. Ganeriwal, and M. B. Srivastava, "Aggregation in sensor networks: An energy-accuracy trade-off," in Proc. of IEEE International Workshop on Sensor Network Protocols and Applications (SNPA'03), Anchorage, AK, pp. 128-138, May 2003.

[9] E. J. Duarte-Melo, M. Liu, and A. Misra, "A modeling framework for computing lifetime and information capacity in wireless sensor networks," in Proc. of 2nd WiOpt:Modeling and Optimization in Mobile, Ad Hoc and Wireless Networks, Cambridge, UK, pp. 45-52,Mar. 2004.

[10]Sourabh Jain, Praveen Kaushik, Jyoti Singhai, "Energy efficient maximum lifetime routing for swireless sensor", Department of Computer Sciene and Engineering Maulana Azad National Institute of Technology, Bhopal, pp. 2122-2129,January2012